\documentclass[onecolumn, aps, prd, groupadress, nofootinbib, superscriptaddress, showkeys, showpacs]{revtex4-1}

\usepackage{amsmath, amsfonts, amssymb, mathrsfs}
\usepackage{amsthm}
\usepackage{tensor}
\usepackage{multirow}
\usepackage{bbm} 
\usepackage[geometry]{ifsym}

\usepackage{cmbright}

\usepackage{graphicx}
\usepackage{float}
\usepackage[font=small]{caption}

\usepackage[hyperindex]{hyperref}
\hypersetup{breaklinks=true, hyperfootnotes=true, pagecolor=white, colorlinks=true}

\usepackage[all]{hypcap}

\makeatletter
\DeclareRobustCommand*{\bfseries}{%
  \not@math@alphabet\bfseries\mathbf
  \fontseries\bfdefault\selectfont
  \boldmath
}
\makeatother

\renewcommand{\nabla}{\!\mathrel{\raisebox{.15em}{%
           \reflectbox{\rotatebox[origin=c]{180}{$\triangle$}}}}\!\!} 
 
\newcommand{\D}{\mathit \Delta} 
\newcommand{\spin}{\mathit \Gamma}

\newcommand{\g}{g_{s}}
\newcommand{\Dslash}{{D\kern-0.63em{/}}}     
\newcommand{\scalf}{a}  
\newcommand{\E}{\tau}

\newcommand{\W}{G}
\newcommand{\G}{G}   

\newcommand{\X}{\chi}

\newcommand{\J}{J}   
\newcommand{\Ji}{J}   
\newcommand{\Wt}{U} 

\newcommand{\Si}{\Sigma}

\newcommand{\Ps}{q}

\newcommand{\B}{B}
\newcommand{\bag}{\beta}

\newcommand{\ca}{\xi}
\newcommand{\cb}{\varepsilon}
\newcommand{\us}{U^2_{\ast}}
\newcommand{\x}{\chi_{\ast}}
\newcommand{\z}{\gamma}

 \newcommand{\Ia}{a}
\newcommand{\Ib}{b}

\newcommand{\Fa}{A}
\newcommand{\Fb}{B}
\newcommand{\Fc}{C}

 \newcommand{\Na}{\alpha}
\newcommand{\Nb}{\beta}

\newcommand{\Nm}{\mu}
\newcommand{\Nn}{\nu}

\newcommand{\Nr}{\rho}
\newcommand{\Ns}{\sigma}

\begin{document}

\title{Gravitational mechanism for baryogenesis in the cosmological QCD phase transition}

\author{V. Antunes}\email[Electronic address:]{antunes@cbpf.br}
 \affiliation{Centro de Estudos Avan\c{c}ados de Cosmologia (CEAC-CBPF), Rua Dr. Xavier Sigaud 150, Urca, CEP 22290-180,
 Rio de Janeiro, RJ, Brazil}
\author{I. Bediaga}\email[Electronic address:]{bediaga@cbpf.br}
 \affiliation{Centro Brasileiro de Pesquisas F\'{i}sicas (CBPF), Rua Dr. Xavier Sigaud 150, Urca, CEP 22290-180,\\
 Rio de Janeiro, RJ, Brazil}{
\author{M. Novello}\email[Electronic address:]{novello@cbpf.br}
 \affiliation{Centro de Estudos Avan\c{c}ados de Cosmologia (CEAC-CBPF), Rua Dr. Xavier Sigaud 150, Urca, CEP 22290-180,
 Rio de Janeiro, RJ, Brazil}

\date{ \today}

\begin{abstract}

 One of the biggest puzzles in modern cosmology is the observed baryon asymmetry in the universe. In current models of baryogenesis gravity plays a secondary role, although the process is believed to have happened in the early universe, under the influence of an intense gravitational field. In the present work we resume Sakharov's original program for baryogenesis and propose a central role for gravity in the process. This is achieved through a non-minimal coupling (NMC) between the gravitational field and both the strong interaction field and the quark fields. When in action, the present mechanism leads to baryon number non-conservation and CP violation. Moreover, the NMC induces reduced effective quark masses, which favours a first order QCD phase transition. As a consequence, a baryon asymmetry can be attained in the transition from the quark epoch to the hadron epoch. 

\end{abstract}


\maketitle

%
%

\section{Introduction \label{intro}}

 
 One of the most puzzling aspects of the interplay between particle physics and cosmology is the observed baryon asymmetry in the universe.
 According to theoretical predictions and recent observations \cite{PDG_NUCLEOSYN, RIOTO_1998, CLINE_2006, DOLGOV_1997}, the measure of this asymmetry is given by
 \begin{equation}
\D \equiv \frac{n_{B} - n_{\bar{B}}}{n_{\gamma}} \approx 6\times 10^{-10},
\label{delta}
\end{equation} 
where $n_{B}$ is the number of baryons per volume, $n_{\bar{B}}$ is the number of atibaryons per volume, and $n_{\gamma}$ is the number of photons per volume in the universe. 
 Subatomic processes in the standard model of particle physics (SM), however, are matter-antimatter symmetric. Fluctuations in the baryon and antibaryon distributions, on the other hand, could only produce an asymmetry far too small to account for the observed baryon asymmetry \cite{RIOTO_1998, CLINE_2006, DOLGOV_1997}.
 
  In a seminal paper \cite{SAKHAROV_1966} Sakharov pointed the way out of this impasse by identifying the necessary conditions for an asymmetric baryogenesis to occur in an earlier stage of the expanding universe, {\it viz.} it is required:
\begin{enumerate}
\item[i)] Baryon number ($B$) non conservation;
\item[ii)] Charge conjugation (C) and combined charge conjugation and parity transformation (CP) symmetries violation;
\item[iii)] Departure from thermal equilibrium. 
\end{enumerate}
Although (iii) was not clearly stated in \cite{SAKHAROV_1966}, conditions (i)-(iii) are now widely known as Sakharov's conditions for baryogenesis. They are not sufficient, however, to ensure an asymmetric baryogenesis, and it is not clear what mechanism was responsible for the baryon asymmetry in our universe. Sakharov went further in \cite{SAKHAROV_1966} and proposed the first explanation for the observed baryon asymmetry based on the assumption that the universe is globally CPT symmetric, that is symmetric under the combined charge conjugation (C), parity transformation (spatial reflection) (P), and time reversal (T) operations, with respect to a singular cosmological bounce. Departure from thermal equilibrium was ensured by the pre-bounce synthesis and subsequent post-bounce decay of hypothetical maximally massive particles (maximons)\footnote{Maximons (or friedmons) are hypothetical particle-like gravitating systems, proposed by Markov \cite{MARKOV_1965, MARKOV_1967, MARKOV_1971, MARKOV_1973}, with mass close to the Planck mass $M_P = \sqrt{\hslash c/G}\sim 10^{-5}$ g. They can also be thought of as elementary black holes.}. As a result an antibaryon excess before the bounce would be exactly converted into the baryon excess of the present-day universe. Clearly, in Sakharov's proposal the baryon asymmetry before the bounce is nothing but an initial condition. Implicit in Sakharov's argumentation, however, is the idea that gravity must have played a crucial role in the asymmetric baryogenesis, since this process must have happened in a much denser and hotter universe, where the gravitational field was much stronger than it is today. This idea has been generally underappreciated since then, and gravity has been downgraded to a mere secondary agent in baryogenesis. 
  
 There is now considerable evidence that the universe was indeed much denser and hotter in the past, with temperatures exceeding $T = 100$ GeV ($k_B = 1$) \cite{PDG_BIG_BANG}. According to the standard cosmological model (SCM), baryogenesis must have happened at some time before the radiation epoch of cosmic history, and shortly after the inflationary epoch. The reason behind this is that any asymmetry in the quark/antiquark ratio before inflation would by diluted to a negligible value due to entropy production during the reheating \cite{OLIVE_1990, ALLAHVERDI_2006}. Within the SM, non-perturbative solutions (sphalerons) could lead to an anomalous $\B + L$ violation, $L$ being the lepton number, resulting from quantum corrections at temperatures of the order of the Electroweak (EW) transition $T\approx 100$ GeV, while preserving $\B - L$ \cite{KUZMIN_RUBAKHV_SHAPOSHNIKOV_1985}. 
The only natural source of CP violation in the SM, however, is related to the occurrence of a phase in Cabibbo-Kobayashi-Maskawa mass matrix in the standard EW sector, which is too small to account for the observed baryon asymmetry. Moreover, the observed Higgs boson mass ($\approx 125$ GeV) implies that the EW transition is not a phase transition of any kind, but a smooth crossover \cite{IGNATIUS_1993, VEGA_2001,STRAUMANN_2004, BVS_2006}. The need of new sources of CP violation and departure from thermal equilibrium has led to the proposal of a number of mechanisms for baryogenesis which rely on SM extensions \cite{DIMOPOULOS_1978,YOSHIMURA_1978, LANGACKER_1981, FUKUGITA_YANAGIDA_1986}. 

 It is believed that the dominant component of the cosmic fluid when temperatures in the universe were as high as 1 GeV was made of deconfined quarks and gluons in a quark-gluon plasma (QGP) state \cite{COLLINS_PERRY_1975, OLIVE_1991, YAGI_2005, MUKHANOV_2005}.  In contrast with the low-energy regime of quantum chromodynamics (QCD), where quarks and gluons are permanently confined to distances of the order of the hadronic radius $r_h \sim 10^{-17}$ cm (colour confinement), in the QGP phase both quarks and gluons can move almost freely through the plasma as a consequence of the asymptotic freedom. Theoretical arguments indicate that the transition from the QGP phase to the hadron gas (HG) phase is expected at a critical temperature $T_c \approx 150$ MeV  \cite{SATZ_2012}. This fact is supported by Lattice QCD (LQCD) simulations \cite{BAZAVOV_2009, BORSANYI_2010}, and also indicated by experimental evidences at the RHIC \cite{MARTINEZ_2013}. Although the exact nature of the transition is not yet fully clear, LQCD results strongly indicate that it depends on the number of quark flavours and the quark masses. For three flavours a strong first order phase transition is predicted in the zero mass limit. When the physical masses of the lightest quarks (up, down ans strange) are taken into account, on the other hand, the predicted quark-hadron transition is not a true phase transition, but rather a sharp crossover \cite{HOTQCD_2014}. 
Finally, the lack of experimental evidence of CP violation in QCD, a puzzling aspect of the theory of strong interactions, forces the natural CP violating terms in the QCD Lagrangian to be fine tuned (strong CP problem) \cite{BIGI_SANDA_2009}. As a consequence no CP violation in the QCD sector is expected in the early expanding universe. 
 
 Regardless of the specific mechanism and the exact epoch, baryogenesis must have taken place in a very dense universe, where gravitational effects cannot be completely neglected. 
 In the present work we resume Sakharov's ``cosmo-physical'' program for baryogenesis and propose to restore the place of gravity at the heart of the process. To achieve that we employ a mechanism which explicitly involves gravity through the non-minimal coupling (NMC) between the gravitational field and the strong interaction field. NMC occurs naturally in field theory, with examples such as $\phi \eta^{\Nm\Nn\Nr\Ns}F_{\Nm\Nn}F_{\Nr\Ns}/2$, which gives a good phenomenological description of the interaction between the neutral pion $\pi^0$ and the electromagnetic field, and also the magnetic dipole interaction $\bar{\psi}\sigma^{\Nm\Nn}\psi F_{\Nm\Nn}$ \cite{JENKINS_2013}. In gravitation, the NMC principle has been widely investigated as a source of non-singular cosmological models \cite{NOVELLO_SALIM_1979, NOVELLO_BERGLIAFFA_2008}, and also in the context of cosmic inflation, where interaction Lagrangians with the form $R\phi^2$ are studied \cite{SALOPEK_BOND_BARDEEN_1989,FAKIR_UNRUH_1990,KAISER_1995,KOMATSU_FUTAMASE_1999,BEZRUKOV_SHAPOSHNIKOV_2008,BARVINSKY_KAMENSHCHIK_STAROBINSKY_2008}. Here we will consider the simplest NMC interaction term between gluon fields and the gravitational field involving the curvature scalar
 \begin{equation*}
 R\,\mbox{tr}\big(\W_{\Nm}\W^{\Nm}\big).
 \end{equation*}
 A direct consequence of this is that colour charge conservation is explicitly violated at the classical level. The fact that Newton's constant is not present in the NMC interaction term is crucial here, since it implies that the difference between the intrinsic energy scale of these theories is not relevant to this process. Moreover, in order to be consistent with colour charge non-conservation, the NMC between the gravitational field and the strong interaction field demands a modification of the Dirac Lagrangian. For that purpose we will also consider the simplest NMC interaction term between quark fields and the gravitational field which explicitly violates parity and dynamically violates charge conservation
 \begin{equation*}
 R\bar{\Ps}\gamma_5\Ps.
 \end{equation*}
The resulting modified quark Lagrangian leads to gravitationally induced effective quark masses which can only be smaller or equal to the bare masses. Although hermiticity is violated, in the cosmological scenario associated with a quark-hadron transition the symmetries of the Lagrangian allow the Hamiltonian to have a real spectrum of eigenvalues, provided that the squared effective quark masses are positive. Moreover, the reduced quark masses which result from the present mechanism can lead to a first order quark-hadron phase transition. Since the gravitational field generated by the cosmic fluid in the quark epoch cannot produce effects in distances of the order of the hadronic radius $r_h \sim 10^{-17}$ cm, only deconfined quarks and gluons can be affected by the mechanism in a cosmological scenario. As soon as quarks and gluons confinement is triggered the present mechanism is switched off, and the baryon asymmetry freezes out. For the sake of simplicity, we will employ the bag model's equation of state for the QGP to estimate the possible effects of the NMC in a cosmological quark-hadron phase transition. 
 
 It has been argued that the transition from the QGP phase to the hadronic phase, the last ``phase transition'' to take place in the universe, was probably not of great relevance for the subsequent cosmic evolution \cite{MUKHANOV_2005}. We intend to show in what follows that the exact opposite may have been the case.

\section{NMC and baryon number violation in the universe}

 \subsection{Charge conservation in the classical Einstein--Yang-Mills--Dirac system}

To fix the notation and recall some basic facts, we consider first the QCD Lagrangian in curved space-time, that is the Yang-Mills Lagrangian for the $SU(N_c)$ group with $N_{c}=3$ (colours) \cite{PDG_QCD}
\begin{widetext}
\begin{equation}
\mathcal{L}_{EYM} = \frac{1}{2\kappa}R -\frac{1}{4}\tensor{\G}{^{\Fa}_{\Nm\Nn}}\tensor{\G}{_{\Fa}^{\Nm\Nn}}
+ \sum_{\Ps} \bigg[ \frac{i}{2}\big( \bar{\Ps}\gamma^{\Nm}D_{\Nm}\Ps - D_{\Nm}\bar{\Ps}\gamma^{\Nm}\Ps \big) - m_{\Ps}\bar{\Ps}\Ps \bigg] ,
\label{qcd_lagrangian}
\end{equation}
\end{widetext}
where $R$ is the curvature scalar, $\kappa$ is the gravitational constant, we have adopted the natural units $k_B = \hslash = c =1$, 
the tensor
\begin{equation}
\tensor{\G}{^{\Fa}_{\Nm\Nn}}(x) =\,\ \nabla_{\Nm}\tensor{\W}{^{\Fa}_{\Nn}} - \nabla_{\Nn}\tensor{\W}{^{\Fa}_{\Nm}}
-i\g[\tensor{\W}{_{\Nm}},\tensor{\W}{_{\Nn}}]^{\Fa}
\label{qcd_field_strength}
\end{equation}
is the field strength, $\nabla_{\Nm}$ is the Riemannian covariant derivative, $\g$ is the bare coupling constant, the gauge fields 
$\tensor{\W}{_{\Nm}}(x) = \tensor{\W}{^{\Fa}_{\Nm}}(x)\lambda_{\Fa}/2$ represent the gluons, the indices $\Fa,\Fb,\Fc,\cdots$ run through the $N_c^2 -1 = 8$ gluon fields, $\lambda_{\Fa}/2$ being the $SU(N_{c})$ Lie algebra generators satisfying $[\lambda_{\Fb},\lambda_{\Fc}] = 2i\tensor{f}{^{\Fa}_{\Fb\Fc}}\lambda_{\Fa}$, where $\tensor{f}{^{\Fa}_{\Fb\Fc}}$ are the structure constants of $SU(N_c)$. Summation is assumed for repeated indices. The spinor fields $\Ps(x) = (\tensor{\Ps}{_{\Ia}}(x))$ represent quarks, where the indices $\Ia,\Ib,\cdots$ run through the $N_{c}=3$ colours, $r,g,b$. Summation in (\ref{qcd_lagrangian}) is with respect to the $N_f$ different quark flavours $\Ps=u,d,s,\cdots$, and $m_{\Ps}$ is the mass of the $\Ps$-flavoured quark. The conjugate spinor is defined as $\bar{\Ps} = \Ps^{\dagger}\gamma^{0}$. Here the Dirac matrices $\gamma_{\Nm}$ obey the relation $\gamma_{\Nm}\gamma_{\Nn} + \gamma_{\Nn}\gamma_{\Nm} = 2g_{\Nm\Nn}$, where $g_{\Nm\Nn}$ are the components of the semi-Riemannian metric with signature $(+---)$. The gauge derivative for fermions is defined as
\begin{equation}
D_{\mu}\Ps \equiv  
\left( \,\nabla_{\Nm} - \spin_{\Nm} - i\frac{\g}{2}\tensor{\W}{^{\Fa}_{\Nm}}\lambda_{\Fa} \right) \Ps,
\label{gauge_derivative_spinors}
\end{equation} 
where $\spin_{\Nm}$ is the spin connection \cite{FOCK_1929}, and $\lambda_{\Fa}$ acts over the quark fields in the fundamental representation of the gauge group (Gell-Mann matrices).

Variation of the action defined by the Lagrangian (\ref{qcd_lagrangian}) with respect to the independent field variables produces the field equations
\begin{subequations}
\begin{equation}
R_{\Nm\Nn} - \frac{1}{2}Rg_{\Nm\Nn} = -\kappa \E_{\Nm\Nn}(\W_{\Na}) - \kappa \E_{\Nm\Nn}(\Ps),
\label{gravity_field_eq}
\end{equation}
\begin{equation}
\nabla_{\Nm}\tensor{\G}{^{\Fa\Nm\Nn}} - i\g[\tensor{\W}{_{\Nm}},\tensor{\G}{^{\Nm\Nn}}]^{\Fa}
= \g\tensor{\J}{^{\Fa\Nn}},
\label{qcd_field_eq}
\end{equation}
plus Dirac equation for quark fields
\begin{equation}
i\gamma^{\Nm}D_{\Nm}\Ps - m_{\Ps}\Ps  = 0.
\label{dirac_eq1}
\end{equation}
\end{subequations}
In the equations above $R_{\Nm\Nn}$ are the components of the Ricci tensor, we have defined the the energy-momentum tensor of the gluons fields
\begin{subequations}
\begin{equation}
\E_{\Nm\Nn}(\W_{\Na}) = -\tensor{\G}{_{\Fa}_{\Nm}^{\Na}}\tensor{\G}{^{\Fa}_{\Nn}_{\Na}} + \frac{1}{4}\tensor{\G}{^{\Fa\Na\Nb}}\tensor{\G}{_{\Fa\Na\Nb}}g_{\Nm\Nn},
\label{energy_momentum_tens_gluons}
\end{equation}
the symmetrized energy-momentum tensor of the quark and antiquark fields
\begin{equation}
\E_{\Nm\Nn}(\Ps) = \sum_{\Ps} i\Big[ \bar{\Ps}\gamma_{\Nm}D_{\Nn}\Ps - D_{\Nm}\bar{\Ps}\gamma_{\Nn}\Ps \Big] + (\Nm \leftrightarrow \Nn), \label{energy_momentum_tens_quarks}
\end{equation}
and it was defined the quark's colour current
\begin{equation}
\tensor{\J}{^{\Fa}_{\Nm}}(x) 
= \sum_{\Ps}\bar{\Ps}\gamma_{\Nm}\frac{\lambda^{\Fa}}{2}\Ps.
\label{colour_current}
\end{equation}
\end{subequations}
From equation (\ref{qcd_field_eq}) follows the conservation of the total (quarks + gluons) colour current 
\begin{equation}
\nabla_{\Nn} \left( \tensor{\J}{^{\Fa\Nn}} + i[\tensor{\W}{_{\Nm}},\tensor{\G}{^{\Nm\Nn}}]^{\Fa}\right) = 0.
\label{colour_current_cons}
\end{equation}
The other conserved Noether currents following from symmetries of the QCD Lagrangian are the 
isospin, related to the global $SU(2)$-symmetry, the chiral current, related to the global axial $U(1)$-symmetry, and the
quark number current 
\begin{equation}
\tensor{\Ji}{_{\Nm}}(x)
= \sum_{\Ps}\bar{\Ps}\gamma_{\Nm}\Ps,
\label{quark_number_current}
\end{equation}
related to the global vectorial $U(1)$-symmetry, the conservation of which is expressed by the equation
\begin{equation}
\nabla_{\Nm}\tensor{\Ji}{^{\Nm}} = 0.
\end{equation}
Integration of $\tensor{\Ji}{_{0}}(x)$ over a spatial section ${}^3\Si(t)$ of space-time gives a conserved charge which, up to a normalization factor $1/3$, corresponds to the baryon number
\begin{equation}
\B \equiv \frac{1}{3}\sum_{\Ps} \int \Ps^{\dagger}(x)\Ps(x)\, d^3 \Si_0 .
\label{baryon_number}
\end{equation}
$\B=1/3$ for a single quark, $\B=-1/3$ for a single antiquark, and $\B=0$ for leptons. Accordingly, one has $\B= 1$ for baryons (three quarks), $\B=-1$ for antibaryons (three antiquarks), and $\B=0$ for mesons (one quark plus one antiquark). More generally
\begin{equation}
\B = N_{B} - N_{\bar{B}},
\end{equation}
where $N_{B}$ is the number of baryons and $N_{\bar{B}}$ the number of antibaryons.

 \subsection{Non-minimally coupled Einstein--Yang-Mills--Dirac system\label{nmc}}

We now consider the NMC extension of the QCD Lagrangian in curved space-time (\ref{qcd_lagrangian}) which results from the inclusion of an additional term
\begin{equation}
\mathcal{L}_{NMC} = -\ca R\tensor{\W}{^{\Fa}_{\Nm}}\tensor{\W}{_{\Fa}^{\Nm}}  +  R\sum_{\Ps} \cb_{\Ps}\bar{\Ps}\gamma_5\Ps,
\label{qcd_lagrangian_nmc}
\end{equation}
where $\gamma^5 \equiv i\gamma^0\gamma^1\gamma^2\gamma^3 = \gamma_5$. Here $\ca$ is an adimensional constant, while the constants $\cb_{\Ps}$, one for each quark flavour $(\Ps = u,d,s,\cdots)$, have dimension of length.
We would like to stress the fact that the coupling constants $\ca$ and $\cb_{\Ps}$ does not involve Newton's gravitational constant. Variation of the action defined by the sum of Lagrangians (\ref{qcd_lagrangian}) and (\ref{qcd_lagrangian_nmc}), $\mathcal{L}_{EYM} + \mathcal{L}_{NMC}$, with respect to the field variables produces the field equations
\begin{widetext}
\begin{subequations}
\begin{align}
\bigg( \frac{1}{\kappa} - \ca\W^2 + &\,\X \bigg)\left( R_{\Nm\Nn} - \frac{1}{2}Rg_{\Nm\Nn} \right) + \ca g_{\Nm\Nn}\nabla^{\Ns}\nabla_{\Ns} \W^{2} - \ca \nabla_{\Nm}\nabla_{\Nn}\W^2 - \ca R\tensor{\W}{^{\Fa}_{\Nm}}\tensor{\W}{_{\Fa}_{\Nn}}
\nonumber\\ & - g_{\Nm\Nn}\nabla^{\Ns}\nabla_{\Ns}  \X\, + \nabla_{\Nm}\nabla_{\Nn}\X
= - \E_{\Nm\Nn}(\W_{\Na}) - \E_{\Nm\Nn}(\Ps),
\label{gravity_field_eq_nmc}
\end{align}
\begin{equation}
\nabla_{\Nm}\tensor{\G}{^{\Fa\Nm\Nn}} - 2\ca R\tensor{\W}{^{\Fa\Nn}} = \g\left( \tensor{\J}{^{\Fa\Nn}} + i[\tensor{\W}{_{\Nm}},\tensor{\G}{^{\Nm\Nn}}]^{\Fa}\right),
\label{qcd_field_eq_cst}
\end{equation}
\end{subequations}
\end{widetext}
where we have defined $\W^2(x) \equiv \tensor{\W}{_{\Fa}_{\Nm}}\tensor{\W}{^{\Fa\Nm}}$, and
\begin{equation}
\X(x) \equiv \sum_{\Ps} \cb_{\Ps}\bar{\Ps}\gamma_5\Ps. \label{q_pseudoscal}
\end{equation}
The energy-momentum tensor for the gluon fields and quark fields have the same forms (\ref{energy_momentum_tens_gluons}) and (\ref{energy_momentum_tens_quarks}), respectively, as before. The Dirac equation, on its turn, now becomes
\begin{subequations}
\begin{equation}
i\gamma^{\Nm}D_{\Nm}\Ps - \big( m_{\Ps} - \cb_{\Ps}R\gamma_5 \big) \Ps = 0.
\label{dirac_nmc_eq1}
\end{equation}
\end{subequations}
As a consequence, we obtain
\begin{equation}
\nabla_{\Nm}\J^{\Nm}  
= 2i R\sum_{\Ps} \cb_{\Ps}\bar{\Ps}\gamma_5\Ps = 2i R\X \neq 0. \label{quark_current_cons_violation_eq}
\end{equation}
On the other hand, according to equation (\ref{qcd_field_eq_cst}), charge conservation is now violated by the non-minimal coupling term
\begin{equation}
\nabla_{\Nn} \left( \tensor{\J}{^{\Fa\Nn}} + i[\tensor{\W}{_{\Nm}},\tensor{\G}{^{\Nm\Nn}}]^{\Fa}\right) = -\frac{2\ca}{\g} \nabla_{\Nn}\left( R\tensor{\W}{^{\Fa\Nn}} \right).
\label{current_cons_violation_eq}
\end{equation}
The NMC term in Lagrangian (\ref{qcd_lagrangian_nmc}) does not alter the form of the energy-momentum tensor of the quark and antiquark fields (\ref{energy_momentum_tens_quarks}). The non-hermiticity of the fermion Lagrangian in general implies the occurrence of non-physical states. This is not the case, however, if the spectrum of eigenvalues of the corresponding Hamiltonian is real. This condition is fulfilled if the Hamiltonian is symmetric under the combined parity transformation (P) and time reversal (T) \cite{BENDER_BOETTCHER_1998, BENDER_2007, MANNHEIM_2015, MANNHEIM_2016}. Therefore we demand that the fermionic sector in the extended Lagrangian (\ref{qcd_lagrangian})+(\ref{qcd_lagrangian_nmc}) leads to a PT-symmetric Hamiltonian. Ignoring, for simplicity, the interaction with the gauge field, iteration of equation (\ref{dirac_nmc_eq1}) yields the analogue of the Klein-Gordon equation
\begin{equation}
\big( \Box 
+ m_{\Ps}^2 - \cb_{\Ps}^2R^2 - i\cb_{\Ps} \gamma^{\Nm}\gamma_5\nabla_{\Nm}R \big) \Ps_{\Ia} = 0, \label{kge}
\end{equation}
where $\Box = \ \nabla_{\Nm}\!\nabla^{\Nm} - R/4$ is the d'Alembertian operator in curved space-time \cite{SCHRODINGER_1932}. In the cosmological scenario that will be discussed in the sequel the contribution from the last term in equation (\ref{kge}) either vanishes or can be ignored, and the regime of unbroken PT-symmetry \cite{BENDER_2007, CASTRO_2011, RODINOV_2013} is defined by the condition\footnote{This fact may imply that each term $\cb_{\Ps}R$ should have the same origin as the mass $m_{\Ps}$ of the corresponding quark. Mechanisms capable of generating mass from non-minimal coupling with the gravitational field  were proposed in \cite{NOVELLO_2011a,NOVELLO_2012}.}
\begin{equation}
 m_{\Ps}^2 \geq \cb_{\Ps}^2R^2. \label{pt_sym}
\end{equation}

 \subsection{Quark-hadron transition\label{qgp}}

   According to perturbative QCD, quarks and gluons become deconfined for temperatures much larger than the QCD scale parameter 100 MeV $\lesssim\Lambda_{QCD} \lesssim$ 300 MeV. This is a consequence of the fact that the effective coupling constant, given to one-loop order by \cite{COLLINS_PERRY_1975}
\begin{equation}
\g^2(T) \simeq \frac{1}{\log (T/\Lambda_{QCD})},
\end{equation} 
becomes significantly small in this regime \cite{KOGUT_STEPHANOV_2004, LETESSIER_RAFELSKY_2004, YAGI_2005, SARKAR_2010, SATZ_2012}.

Qualitatively this transition can be understood by resorting to a bag model \cite{CHODOS_1974, GREINER_QCD, LETESSIER_RAFELSKY_2004}. In these phenomenological models of the strong interaction, in the hadronic phase ($T < T_c$) 
quarks and gluons are assumed to be confined inside bounded spatial regions (bags) with radius $r_h \sim 10^{-17}$ cm.  Boundary condition are imposed so that the quark current, the chromo-electric field, and the chromo-magnetic field all vanish at the bag's boundary. Light quarks and high momentum (hard) gluons are assumed to move freely inside the bag, while the non-perturbative contribution from low momentum (soft) gluons is given by a constant energy density which exerts a negative pressure preventing quarks from escaping the bag, as expressed by the energy-momentum tensor for the confined fields
\begin{equation}
\E_{\Nm\Nn}(\Ps,\W_{\Na}) = \Big( \E_{\Nm\Nn}(\Ps) + \E^{(0)}_{\Nm\Nn}(\W_{\Na}) + \bag g_{\Nm\Nn} \Big)\theta(r_h-r),
\end{equation}
where $\E^{(0)}_{\Nm\Nn}(\W_{\Na})$ is energy-momentum tensor for the perturbative gluon contribution, and $\theta(r_h-r)$ denotes the step function. Macroscopically, this hadronic phase is described as an ideal gas composed mostly of ultrarelativistic pions, $\pi^0$ ($u\bar{u}$ or $d\bar{d}$), $\pi^+$ $(u\bar{d})$, and $\pi^-$ $(d\bar{u})$, plus a rare fraction of nucleons carrying the cosmic baryon number, with equation of state 
\begin{equation}
\rho_{h}(T) = 3p_{h}(T) = \sigma_{h} T^4,
\label{hg_eos}
\end{equation}
where $\sigma_{h} = (\pi^2/30) \left( d_{h} + d_{\gamma+\ell} \right) \simeq 19$ is the Steffan-Boltzmann constant for a hadron gas (HG), $d_h \simeq 5$ being hadronic degrees of freedom, and $d_{\gamma+\ell} = 14.25$ is the photon + leptons degrees of freedom. As the temperature approaches the critical value $T_{c}$ from below, the bag radius grows and individual bags eventually overlap, the bag pressure now being equal to the bag constant $\bag$ everywhere. 

 Alternatively, deconfinement can be characterized by the generalized form of the Cornell potential describing the interaction between static colour charges \cite{SABZEVARI_2002, DORING_2007, SATZ_2012}
\begin{equation}
U(r,T) = -\frac{C(T)}{r}e^{-m_Dr} + K r \left( \frac{1 - e^{-m_Dr}}{m_Dr} \right), \label{qq_potential}
\end{equation}
where $C(T)$ is a temperature-dependent parameter, $K$ is the string tension, $r$ is the distance between two colour charged particles, and $m_D(T) = 1/\lambda_D$ is the Debye screening mass given, to leading perturbative order and for a vanishing chemical potential, by \cite{LE_BELLAC_1996, LETESSIER_RAFELSKY_2004}
\begin{equation}
m_D(T) = \left( \frac{N_c}{3} + \frac{N_f}{6} \right)^{1/2} \g(T)T. \label{debye_mass}
\end{equation}

Macroscopically, deconfinement implies that for temperatures higher than a critical temperature $T_c$ nuclear matter assumes the form of a quark-gluon plasma (QGP), that is a perfect fluid composed of unconfined quarks and gluons \cite{COLLINS_PERRY_1975}. Again, chemical potentials can be ignored to a good approximation. The simplest QGP description is given by the bag equation of state \cite{KOGUT_STEPHANOV_2004, LETESSIER_RAFELSKY_2004, YAGI_2005, SARKAR_2010, SATZ_2012} 
\begin{subequations}\label{pre_eos}
\begin{equation}
\rho_{qg}(T) = \sigma_{qg} T^4 + \bag, 
\end{equation}
\begin{equation}
p_{qg}(T) = \frac{1}{3}\sigma_{qg} T^4 - \bag,  
\end{equation}
\end{subequations}
in which $\bag$ is the bag constant, and $\sigma_{qg}= (\pi^2/30)\left( d_{g} + 7d_{\Ps}/4 + d_{\gamma+\ell} \right)$ is the Steffan-Boltzmann constant for the QGP, $d_{g} = 2_s\times (N_c^2 -1)(1 - 15\g^2/8\pi^2)$ and $d_{\Ps} = 2_s\times N_f \times N_c(1 - 50\g^2/84\pi^2)$ being degeneracy factors for the gluons and quarks, respectively. In what follows it will be considered $N_f = 2.5$ as a qualitative account of semi-massive strange quarks \cite{LETESSIER_RAFELSKY_2004}. The bag equation of state (\ref{pre_eos}) predicts a strong first order phase transition, with an abrupt change in the energy density $\rho(T)$, pressure $p(T)$, and entropy $s(T)$, and a significant decrease in the number of degrees of freedom of the system as the temperature drops below $T_c$ \cite{BAZAVOV_2009, BORSANYI_2010}. 
Since $p_{qg}(T_c) = p_{h}(T_c)$ during the phase transition, the critical density in this phenomenological model can be readily obtained from equations (\ref{hg_eos}) and (\ref{pre_eos}). 
For a bag constant $\bag^{1/4} \approx 190\,\mbox{MeV}$ one obtains the critical temperature $T_c \approx 160$ MeV for this phenomenological model.

 In the strong interacting regime $T_c < T \lesssim 3T_c$ gluon interactions, although not confining, are still large enough to produce non-perturbative effects. The form of the QGP equation of state in this regime is not known exactly. LQCD simulations, however, indicate that the nature of the transition from the quark-gluon phase to the hadronic phase, and also the critical temperature of the transition, are strongly dependent on the quark masses and the number of quark flavours \cite{BAZAVOV_2009, BORSANYI_2010, SATZ_2012, HOTQCD_2014}. For three quark flavours a strong first order phase transition is predicted in the zero mass limit. On the other hand, when the physical masses of the three lightest quarks are considered ($m_u \approx 2.3$ MeV for the quark up, $m_d \approx 4.8$ MeV for the quark down, and $m_s \approx 95$ MeV for the quark strange), instead of a true phase transition, LQCD simulations predict a sharp crossover at a critical temperature $T_c \approx 175$ MeV \cite{VEGA_2001, BAZAVOV_2009, BORSANYI_2010, BVS_2006, HOTQCD_2014}.

\subsection{Cosmological quark-hadron transition and asymmetric baryogenesis}

 We now proceed to the implementation of a gravitationally induced baryogenesis mechanism in the quark epoch. Being a very dense phase of the early expanding universe, assumed to be spatially homogeneous and isotropic, the spatial curvature in the quark epoch can be ignored and the space-time metric assume the flat Friedman-Robertson-Walker (FRW) form
\begin{equation}
ds^2 = dt^2 - \scalf^2(t)\big( dr^2 + r^2d\vartheta^2 + r^2\sin^2\vartheta d\varphi^2 \big),
\label{flrw_metric}
\end{equation}
where $\scalf(t)$ is the scale factor. The basic assumptions we make here are first that the strong interaction field is non-minimally coupled to the gravitational field according to the Lagrangian (\ref{qcd_lagrangian_nmc}), and second that the non-minimal coupling does not lead to significant modifications in the QGP sate (at least near the quark-hadron transition). This is ensured by assuming that for $T\geq T_c$ the following relation holds
\begin{equation}
\big|\ca R(T)\big| \ll m_D^2(T).
\label{nmc_condition}
\end{equation}
On the other hand, it could still be large enough to produce effects in the field equation (\ref{qcd_field_eq_cst}).

 Hadronic matter can be described by means of the energy-momentum tensor
\begin{equation}
\E_{\Nm\Nn} = \big\langle \E_{\Nm\Nn}(\W_{\Na}) + \E_{\Nm\Nn}(\Ps) \big\rangle = (\rho + p)V_{\Nm}V_{\Nn} - pg_{\Nm\Nn},
\label{qgp_energy_momentum_tens}
\end{equation} 
where the brackets $\langle\ \ \rangle$ denote an averaging process, $V^{\Nm} = \delta^{\Nm}_0$ are the components of the four-velocity of a comoving observer. In the QGP phase ($T>T_c$) the energy density $\rho(T)$ and pressure $p(T)$ are given by equations (\ref{pre_eos}), while for the HG phase ($T<T_c$) they are given by (\ref{hg_eos}).

 According to the general description of the QGP state outlined in the previous section, for $T$ close to $T_c$ the potential (\ref{qq_potential}) has almost everywhere the form
\begin{equation}
U(T) \simeq \frac{K}{m_D(T)}. \label{qq_potential_lt}
\end{equation}
The break of gauge invariance of the QCD Lagrangian induced by the NMC allows us to take the following components for the gluon fields
\begin{equation}
\tensor{\W}{^{\Fa}_{0}} = \Wt(T),  \ \ \tensor{\W}{^{\Fa}_{i}} = 0,  \ \ \mbox{for all}\ \Fa = 1,\cdots , 8,  \label{qq_potential_b}
\end{equation}
Assuming colour independence of the quark fields in the QGP, {\it i.e.} $\Ps_{\Ia} = \Ps$ for all $\Ia= r,g,b$, the field equations (\ref{gravity_field_eq_nmc})-(\ref{qcd_field_eq_cst}) finally reduce to
\begin{widetext}
\begin{subequations}
\begin{equation} 
3\left(  \frac{\dot{\scalf}}{\scalf} \right)^2 =
 \frac{1}{\frac{1}{\kappa} - 8\ca \Wt^2 + \X}\big( \rho - 8\ca R\Wt^2 \big),
\label{gravity_field_eq_nmc_flrw_00}
\end{equation}
\begin{equation}
2\frac{\ddot{\scalf}}{\scalf} + \left( \frac{\dot{\scalf}}{\scalf} \right)^2 = 
- \frac{1}{\frac{1}{\kappa} - 8\ca \Wt^2 + \X}\Big[ p - 16\ca \big( \Wt\ddot{\Wt} + \dot{\Wt}^2 \big) + \ddot{\X}\Big] ,
\label{gravity_field_eq_nmc_flrw_ii}
\end{equation}
\begin{equation}
2\ca R\Wt +  \g\tensor{\Ji}{_{0}} = 0,
\label{qcd_field_eq_cst_c0}
\end{equation}
\end{subequations}
\end{widetext}
together with a null spatial current $\tensor{\Ji}{_{i}}(x) = 0$, where we have defined $\dot{A} \equiv V^{\Nm}\,{\nabla_{\Nm}A} = {\nabla_{0}}A$. The cosmic evolution equations (\ref{gravity_field_eq_nmc_flrw_00})-(\ref{gravity_field_eq_nmc_flrw_ii}) recover the Friedmann equations for $T< T_{c}$, the cosmic dynamics becoming identical to that described by the SCM. The extrapolation of the past behaviour of the universe from present-time observations remains the same until the quark-hadron phase transition is reached. Moreover, we can suppose that
\begin{equation}
\X \sim \ca\Wt^2. \label{condition_b}
\end{equation}

According to conditions (\ref{nmc_condition}), in the QGP phase ($T>T_c$) the Friedman equation (\ref{gravity_field_eq_nmc_flrw_00}) can be approximated by
\begin{equation}
3\left(  \frac{\dot{\scalf}}{\scalf} \right)^2 \simeq
\kappa \rho. \label{arrox_fried}
\end{equation}
Also, by the same reason, an approximate conservation of matter-energy equation for this regime is obtained
\begin{equation}
\frac{\dot{\scalf}}{\scalf} \simeq -\frac{\dot{\rho}}{3(\rho + p)}.  \label{arrox_conserv}
\end{equation}
From equations (\ref{arrox_fried}) and (\ref{arrox_conserv}), it follows the the energy density of the fluid satisfies
\begin{equation}
-\frac{d\rho}{3\sqrt{\rho}(\rho + p)} \simeq \left( \frac{\kappa}{3} \right)^{1/2}dt.
\label{int_time}
\end{equation}
Employing the bag equation of state (\ref{pre_eos}), equation (\ref{int_time}) yields the following form for scale factor in the QGP phase \cite{LETESSIER_RAFELSKY_2004, YAGI_2005}
\begin{equation}
\scalf(t) \propto (\sinh t)^{1/2}.
\end{equation}
As a consequence, $R \simeq \mbox{const.}$ before the quark-hadron transition, and the last term in the Klein-Gordon equation (\ref{kge}) vanishes until the onset of the transition. This implies that the condition for a Hamiltonian with unbroken PT-symmetry assumes, in this phase, exactly the form (\ref{pt_sym}). As mentioned in the previous section, reduced quark masses can lead to a first order QCD phase transition. Given that, as suggested by LQCD data, the transition is very close to a first order phase transition for the physical bare masses, we can suppose that the present mechanism induces a first order quark-hadron phase transition.

 During the first order phase transition, the cosmic expansion rate ($\sim 10^{-5}$ s) is far greater than the QCD time-scale ($\sim 10^{-23}$ s). As a consequence, the matter-energy content of the universe can be described as a single adiabatically expanding fluid (small supercooling) \cite{CHRISTIANSEN_MADSEN_1996, IGNATIUS_1993, VEGA_2001}, and both the temperature and pressure remain constant during the transition. Accordingly, we can assume that the QGP remains locally colour neutral. We denote by $t_i$ the initial time of the phase transition, and $t_f$ its final time. At the onset of the phase transition HG bubbles form via homogeneous bubble nucleation. The energy density of the cosmic fluid exhibit mixed QGP and HG phases during the transition ($t_i \leq t \leq t_f$), and can be expressed as
\begin{equation}
\rho(t) = f(t){\rho_{qg}}_{\ast} + \big(1-f(t)\big){\rho_{h}}_{\ast}, \label{dens_pht}
\end{equation}
where the energy densities $\rho_{qg}$ and $\rho_{h}$, are given by equations (\ref{pre_eos}) and (\ref{hg_eos}), respectively, and the subscript $\ast$ denotes values at the onset of the phase transition 
\begin{equation}
{\rho_{qg}}_{\ast} \equiv \rho_{qg}(T_c) , \, \ {\rho_{h}}_{\ast} \equiv \rho_{h}(T_c),\, \ {g_s}_{\ast} \equiv \g(T_c),\, \ \mbox{\it etc}.
\end{equation}
Here $f(t)$ denotes the fraction of space occupied by the QGP relative to the fraction occupied by the hadronic matter (pion gas) as a function of the cosmic time, and takes values $0 \leq f(t) \leq 1$. By definition $f(t_i) = 1$ before and at the onset of the transition ($T>T_c$), and $f(t_f)= 0$ at the end of the transition ($T<T_c$). Expression analogous to (\ref{dens_pht}) holds during the phase transition for the pressure $p(t)$ and for the squared inter-coloured charges potential (\ref{qq_potential_lt})
\begin{equation}
\Wt^2(t) = f(t){U^2_{qg}}_{\ast} + \big(1-f(t)\big){U^2_{h}}_{\ast} = \us f(t), \label{potential_pht}
\end{equation}
where we have taken into account the fact that the hadron-hadron interaction is described by a short-range potential \cite{HAGEDORN_1965, BEDIAGA_2000}, {\it i.e.} ${U_{h}}_{\ast}= 0$ at large scale. By a similar reasoning, we have
\begin{equation}
\X(t) = \x f(t).  \label{chiral_curr_pht}
\end{equation}
With the above definitions, the contraction of the gravitational field equation (\ref{gravity_field_eq_nmc}) yields the following approximate expression for the curvature scalar during the phase transition
\begin{equation}
R(t) \simeq \kappa\Big[ 4\bag f(t) + \left( 24\ca \us - 3\x \right) \ddot{f}(t) \Big]. \label{curv_scal_pht}
\end{equation} 
From equation (\ref{current_cons_violation_eq}), the potential (\ref{qq_potential_b}), and again assuming colour independence of the quark fields, we finally obtain
\begin{equation}
b(t) 
= -\frac{2\ca}{3{g_s}_{\ast}}\int_{t_i}^{t} \big( \Wt\dot{R} + R\dot{\Wt} \big) dt', \label{baryon_n_dens}
\end{equation}
where $b(t)= n_B(t) - n_{\bar{B}}(t)$, and $b(t_i) = 0$. 

For simplicity, we will consider here the bag equation of state (\ref{pre_eos}) for the QGP to estimate the possible effects of the NMC over baryogenesis in a cosmological scenario. This choice is in accordance with a first order phase transition scenario. Equation (\ref{int_time}) can again be solved analytically, and yields the following form for the QGP volume fraction function \cite{FMA_1988, YAGI_2005}
\begin{widetext}
\begin{equation}
f(t) = \frac{1}{4(\z -1)}\left\{ \tan^2\left[ \arctan \sqrt{4\z -1} -\frac{3\z}{2\sqrt{\z-1}}\frac{(t-t_i) }{\tau_{\ast}} \right] - 3 \right\}, \label{f}
\end{equation}
\end{widetext}
where $\z$ is a constant factor, and $\tau_{\ast} \equiv \sqrt{3/(4\kappa \bag)}$ is the typical time scale of the quark-hadron transition \cite{LETESSIER_RAFELSKY_2004, YAGI_2005}. In what follows we set $\gamma = 3$, and $t_i = 0$. Using (\ref{f}) in (\ref{potential_pht}), (\ref{chiral_curr_pht}) and (\ref{curv_scal_pht}), we can finally solve equation (\ref{baryon_n_dens}) analytically. 

Figure (Fig. \ref{baryo1}) shows the behaviour of the baryon number density $b(t)$ as a function of the cosmic time obtained by solving equation (\ref{baryon_n_dens}) for different values of the bag constant, $\bag^{1/4} = 170$ MeV (dashed line), $\bag^{1/4} = 190$ MeV (solid line), and $\bag^{1/4} = 220$ MeV (doted line), where we have set ${g_s}_{\ast} = 2.5$, and $K = 0.15\,\mbox{GeV}^2$, and $\ca = 2.6\times 10^{8}$. To these values correspond, respectively, $T_c = 143.4$ MeV, $T_c = 160.3$ MeV and $T_c = 185.6$ MeV for the critical temperatures, $\tau_{\ast} = 48\times 10^{-6}$ s, $\tau_{\ast} = 38.4\times 10^{-6}$ s and $\tau_{\ast} = 28.7 \times 10^{-6}$ for the typical time scales, and $t_f = 3.48 \times 10^{-6}$ s, $t_f = 2.78 \times 10^{-6}$ s, and $t_f = 2.08 \times 10^{-6}$ s for the final transition times.
Figure (\ref{baryo2}) shows the behaviour of the baryon number density $b(t)$ as a function of the cosmic time for different values of the adimensional NMC coupling constant, $\ca = 2.6\times 10^{8}$ (solid line), $\ca = 1.8 \times 10^{8}$ (dashed line), and $\ca = 3.4 \times 10^{8}$ (doted line), where we have set $\bag^{1/4} = 190$ MeV, ${g_s}_{\ast} = 2.5$, and $K = 0.15\,\mbox{GeV}^2$. 
Figure (Fig. \ref{baryo3}) shows the behaviour of the curvature scalar $R(t)$ as a function of the cosmic time during the phase transition for different values of the bag constant, $\bag^{1/4} = 170$ MeV (dashed line), $\bag^{1/4} = 190$ MeV (solid line), and $\bag^{1/4} = 220$ MeV (doted line), where we have set ${g_s}_{\ast} = 2.5$, $K = 0.15\,\mbox{GeV}^2$, and $\ca = 2.6\times 10^{8}$.
Finally, figure (Fig. \ref{baryo4}) shows the behaviour of the curvature scalar $R(t)$ as a function of the cosmic time during the phase transition for different values of the adimensional NMC coupling constant, $\ca = 2.6\times 10^{8}$ (solid line), $\ca = 1.8 \times 10^{8}$ (dashed line), and $\ca = 3.4 \times 10^{8}$ (doted line), where we have assumed $\bag^{1/4} = 190$ MeV, ${g_s}_{\ast} = 2.5$, and $K = 0.15\,\mbox{GeV}^2$.

\begin{figure}[h]
   \centering
  \includegraphics[scale=
0.7 
]{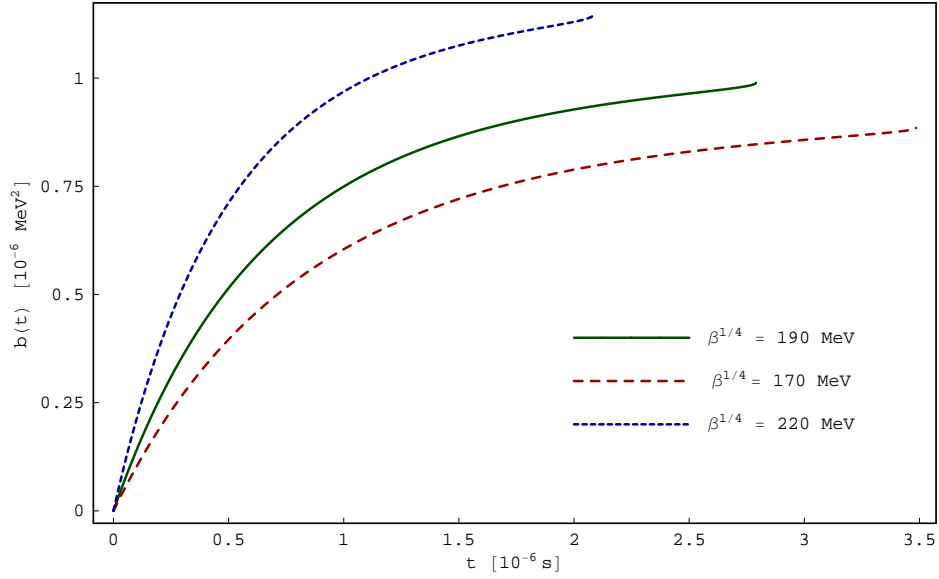} 
   \caption{Plot of the baryon number density $b(t)$ as a function of time during the phase transition for different values of the bag constant, $\bag^{1/4} = 170$ MeV (dashed line), $\bag^{1/4} = 190$ MeV (solid line), and $\bag^{1/4} = 220$ MeV (doted line). We have set $\ca = 2.6\times 10^{8}$, ${g_s}_{\ast} = 2.5$, $t_i = 0$, and $b(0)=0$.} \label{baryo1}
   \end{figure} 
   
\begin{figure}[h]
   \centering
  \includegraphics[scale=
0.7
]{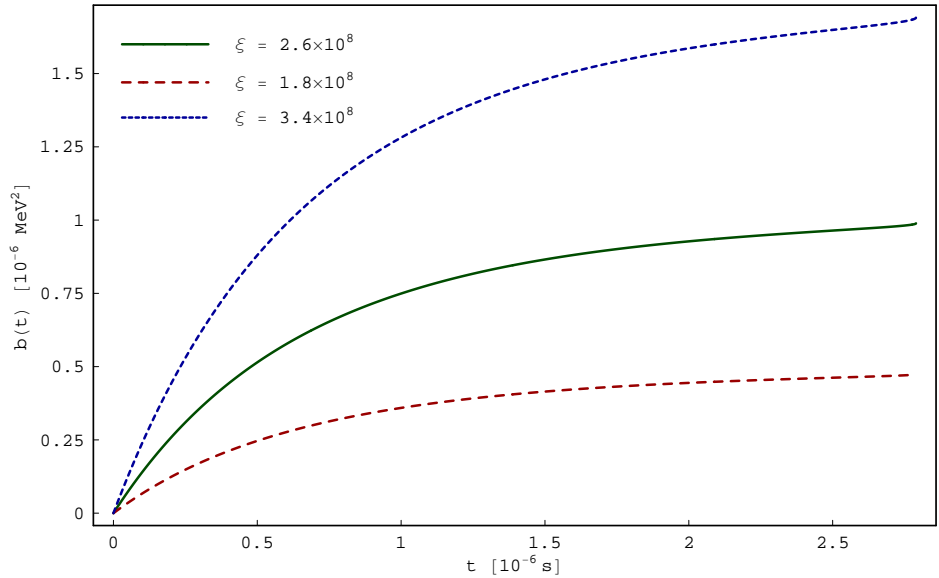} 
   \caption{Plot of the baryon number density $b(t)$ as a function of time during the phase transition for $\ca = 2.6\times 10^{8}$ (solid line), $\ca = 1.8 \times 10^{8}$ (dashed line), and $\ca = 3.4 \times 10^{8}$ (doted line). We have set $\bag^{1/4} = 190$ MeV, ${g_s}_{\ast} = 2.5$, $t_i = 0$, and $b(0)=0$.} \label{baryo2}
   \end{figure} 
   
 \begin{figure}[h]
   \centering
  \includegraphics[scale=
0.7 
]{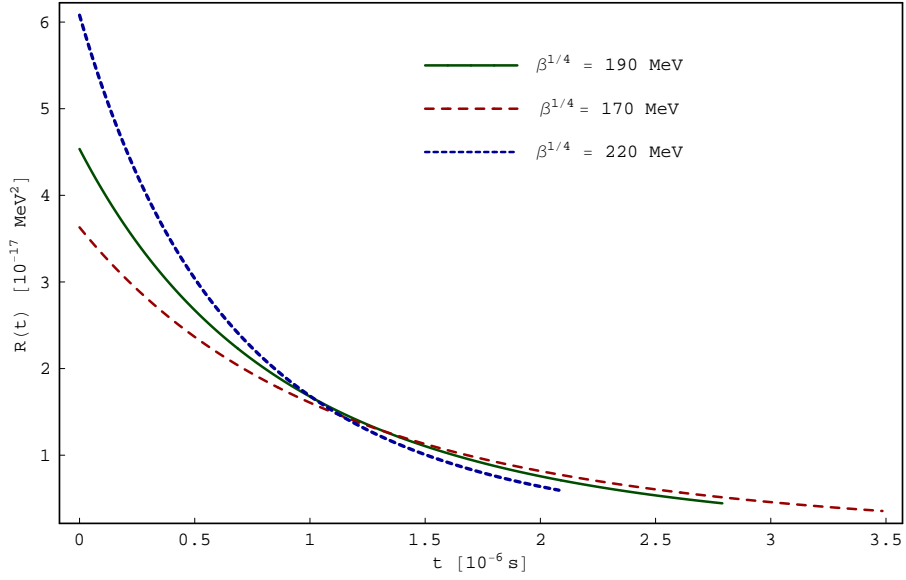} 
   \caption{Plot of the curvature scalar $R(t)$ as a function of time during the phase transition for different values of the bag constant, $\bag^{1/4} = 170$ MeV (dashed line), $\bag^{1/4} = 190$ MeV (solid line), and $\bag^{1/4} = 220$ MeV (doted line). We have set $\ca = 2.6\times 10^{8}$, ${g_s}_{\ast} = 2.5$, $t_i = 0$, and $b(0)=0$.} \label{baryo3}
   \end{figure} 
   
   \begin{figure}[h]
   \centering
  \includegraphics[scale=
0.7 
]{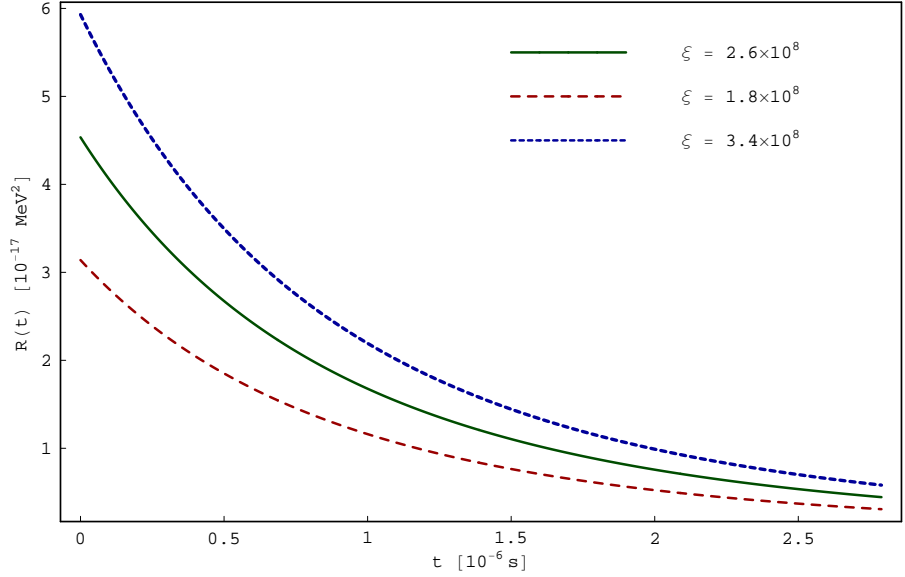} 
   \caption{Plot of the curvature scalar $R(t)$ as a function of time during the phase transition for $\ca = 2.6\times 10^{8}$ (solid line), $\ca = 1.8 \times 10^{8}$ (dashed line), and $\ca = 3.4 \times 10^{8}$ (doted line). We have set $\bag^{1/4} = 190$ MeV, ${g_s}_{\ast} = 2.5$, $t_i = 0$, and $b(0)=0$.} \label{baryo4}
   \end{figure}

  Baryon-antibaryon annihilations freeze at a temperature $T\simeq 20$ MeV, to which corresponds the photon number density
\begin{equation}
n_{\gamma} \simeq \frac{2}{\pi^2}T^3 = 1.6\times 10^3 \ \mbox{MeV}^3.
\end{equation}
For $\ca = 2.6\times 10^{8}$ we obtain $b(t_f) \simeq 9\times 10^{-7}\,\mbox{MeV}^{3}$ at the end of the phase transition. In this case, for $T\leq 20$ MeV corresponds a baryon number density to photon number density ratio
\begin{equation}
\D = \frac{b(t_f)}{n_{\gamma}} \simeq 6 \times 10^{-10},
\end{equation} 
which is compatible with the observed baryon asymmetry today (\ref{delta}).

For values $\ca \sim 10^{8}$ the bound (\ref{nmc_condition}) for $\ca$ is satisfied by a large margin, since in this case ${m^2_{D}}_{\ast}/R_{\ast} \sim 10^{21}$ at the onset of the phase transition. The condition (\ref{pt_sym}), which defines the regime of unbroken PT-symmetry, can also be satisfied by a large margin according to 
\begin{equation}
\cb_{\Ps} \leq \frac{m_{\Ps}}{R},  \ \ \Ps = u,d ,s,
\end{equation}
which reads $\cb_{\Ps} \lesssim 10^{16} \mbox{---} 10^{18}\,\mbox{MeV}^{-1}$ for $\ca \simeq 10^8$, the upper limit $\sim 10^{16}$ applying to the lighter quarks, up and down, and $\sim10^{18}$ to the quark strange. As an example, taking $\cb_{u}= 5\times 10^{16}\,\mbox{MeV}$, $\cb_{d}= 9.7\times 10^{16}\,\mbox{MeV}$, and $\cb_{s}= 2\times 10^{18}\,\mbox{MeV}$ for $\ca = 2.6\times 10^{8}$, $m_u = 2.3$ MeV, $m_d = 4.5$ MeV, and $m_s = 95$ MeV, we obtain the effective quark masses $m^{eff}_q \simeq \sqrt{ m_q^2 - \cb_q^2 R_\ast^2 }$ at the onset of the phase transition
\begin{equation*}
m^{eff}_u \simeq 0.45\ \mbox{MeV},\ \ m^{eff}_d \simeq 0.9\ \mbox{MeV},\ \ m^{eff}_s \simeq 19\ \mbox{MeV}.
\end{equation*}

\section{Conclusion}

In this work we have considered the effects of a non-minimal coupling (NMC) between the gravitational field and both the gluon and quark fields on the asymmetric production of quarks over antiquarks in the early expanding universe. The consequences of the NMC, when the mechanism is in action, are:

\begin{itemize}
\item[(i)] Baryon number ($B$) non conservation;
\item[(ii)] CP violation;
\item[(iii)] Gravitationally induced effective quark masses smaller than the bare physical masses.
\end{itemize}

 Baryon number non conservation is a consequence of colour current non-conservation in a locally colour neutral quark-gluon plasma (QGP). The violation of colour current conservation, however, can only be achieved through the inclusion of an NMC term which violates the hermiticity of the quark Hamiltonian. Accordingly, the gravitationally induced contribution to the effective masses of quarks must be smaller or equal to the bare quark masses in order to preserve PT symmetry, a necessary condition for particle states to be physical. This fact may be relevant for the cosmological dynamics of a QGP dominated universe since, according to lattice QCD simulations, a first order quark-hadron phase transition is predicted for the three quarks flavours in the zero mass limit. The reduced effective quark masses induced by the NMC can therefore lead to a true first order phase transition, instead of a crossover.

Another important fact about the the mechanism discussed here is that it cannot produce effects in ordinary astrophysical environments, such as neutron stars. The reason behind this is that, with the exception of the vicinity of the event horizons of black holes, curvature effects are negligible at distances of the order of the hadronic radius $r_h \sim 10^{-17}$ cm. It has been hypothesized, however, that the nuclear matter at the core of superdense neutron stars could consist of deconfined quarks in a QGP state \cite{IVANENKO_KURDGELAIDZE_1965, IVANENKO_KURDGELAIDZE_1969}. In any case, since neutron star interiors are almost in perfect equilibrium \cite{SHAPIRO_TEUKOLSKY_2004}, no net effect from the present mechanism is expected in such environments. We conclude that at the astrophysical scale the baryon number violation mechanism described here could only be associated with violent processes, such as supernovas/hypernovas or neutron star mergers, with a massive production of gamma rays. 

In a cosmological scenario, since the gravitational field generated by the cosmic fluid in the quark epoch cannot produce effects in distances of the order of the hadronic radius $r_h \sim 10^{-17}$ cm, only unconfined quarks and gluons can be affected by the present mechanism. As soon as quarks and gluons confinement is triggered the present mechanism is switched off and the baryon asymmetry freezes out. For particular values of the adimensional coupling constant appearing in the NMC term between gravity and the strong interaction field, a baryon asymmetry compatible with the observed value is obtained.

\section*{Acknowledgements}

V.A. thanks the support from the Research Support Foundation of the State of Rio de Janeiro (FAPERJ), and
M.N. the support from the Brazilian National Council of Technological and Scientific Development (CNPq).


\begin{thebibliography}{70}
\bibitem{PDG_NUCLEOSYN} B. D. Fields, P. Molaro, and S. Sarkar (Particle Data Group), ``Big-Bang Nucleosynthesis", PDG reviews, \href{http://pdg.lbl.gov}{pdg.lbl.gov} (2015).
\bibitem{RIOTO_1998} A. Rioto, ``Theories of Baryogenesis", Lectures delivered at the Summer School in High Energy Physics and Cosmology, Trieste, Italy, 29 June -17 July 1998; \href{https://arxiv.org/abs/hep-ph/9807454}{arXiv:hep-ph/9807454}.
\bibitem{CLINE_2006} J. M. Cline, ``Baryogenesis", Les Houches Summer School, 86: Particle Physics and Cosmology: the Fabric of Spacetime, 7-11 Aug. 2006; \href{https://arxiv.org/abs/hep-ph/0609145}{arXiv:hep-ph/0609145}.
\bibitem{DOLGOV_1997} A. D. Dolgov, ``Baryogenesis, 30 years after", Lectures given at the 25th ITEP winter school, 18 - 27 Feb 1997, \href{https://arxiv.org/abs/hep-ph/9707419}{arXiv:hep-ph/9707419}.
\bibitem{SAKHAROV_1966} A. D. Sakharov, 
JETP Lett 5, 24–27 (1967). 
\bibitem{MARKOV_1965} M. A. Markov, 
Suppl. Progr. Theor. Phys., extra number, 85 (1965). 
\bibitem{MARKOV_1967} M. A. Markov, 
Soviet. Phys. JETP 24, n. 3, 584 (1967). 
\bibitem{MARKOV_1971} M. A. Markov, ``Cosmology and Elementary Particles", Lecture Notes, International Centre for Theoretical Physics, Trieste, 1971, IC/71/33, parts I and II.
\bibitem{MARKOV_1973} V. I. Man'ko, M. A. Markov, 
Theor. Math. Phys. 17, 1060 (1973). 
\bibitem{PDG_BIG_BANG} K.A. Olive and J. A. Peacock, ``Big-Bang Cosmology", PDG reviews, \href{http://pdg.lbl.gov}{pdg.lbl.gov} (2015).  
\bibitem{OLIVE_1990} K.A. Olive, 
Phys. Rept. 190, 307 (1990).
\bibitem{ALLAHVERDI_2006} R. Allahverdi, R. Brandenberger, F.-Y. Cyr-Racine and A. Mazundar, 
Annu. Rev. Nucl. Part. Sci. 60, 27 (2010); \href{https://arxiv.org/abs/1001.2600}{arXiv:1001.2600 [hep-th]}.
\bibitem{KUZMIN_RUBAKHV_SHAPOSHNIKOV_1985} V. A. Kuzmin, V. A. Rubakov and M. E. Shaposhnikov, 
Phys. Lett. B 155, 36 (1985).
\bibitem{IGNATIUS_1993} J. Ignatius, ``Cosmological Phase Transitions", Ph.D. dissertation, 1993, \href{https://arxiv.org/abs/hep-ph/9312293}{arXiv:hep-ph/9312293}.
\bibitem{VEGA_2001} H. J. De Vega, I. M. Khalatnikov, N. G. S\'{a}nchez (eds.), ``Phase Transitions in the Early Universe", Springer, 2001. 
\bibitem{STRAUMANN_2004} N. Straumann, 
(2004) \href{https://arxiv.org/abs/astro-ph/0409042}{arXiv:astro-ph/0409042}. 
\bibitem{BVS_2006} D. Boyanovsky, H. J. de Vega, and D. J. Schwarz, 
Ann. Rev. Nucl. Part. Sci. 56, 441 (2006); \href{https://arxiv.org/abs/hep-ph/0602002}{arXiv:hep-ph/0602002}.
\bibitem{DIMOPOULOS_1978} S. Dimopoulos and L. Susskind, 
Phys. Rev. D 18, 4500 (1978).
\bibitem{YOSHIMURA_1978} M. Yoshimura, 
Phys. Rev. Lett. 41, 281 (1978).
\bibitem{LANGACKER_1981} P. Langacker, 
Phys. Rept. 72, 185 (1981).
\bibitem{FUKUGITA_YANAGIDA_1986} M. Fukugita and T. Yanagida, 
Phys. Lett. B 174, 45 (1986).
\bibitem{COLLINS_PERRY_1975} J. C. Collins and M. J. Perry, 
Phys. Rev. Lett. 34, 1353 (1975).
\bibitem{OLIVE_1991} K.A. Olive, 
Science 251, 1194 (1991).
\bibitem{YAGI_2005} K. Yagi, T. Hatsuda, and Y. Miake, ``Quark-gluon Plasma: from Big Bang to Little Bang", Cambridge University Press, 2005.
\bibitem{MUKHANOV_2005} V. Mukhanov, ``Physical Principles of Cosmology", Cambridge University Press, 2005. 
\bibitem{SATZ_2012} H. Satz, ``Extreme States of Matter in Strong Interaction Physics", Springer, 2012.
\bibitem{BAZAVOV_2009} A. Bazavov, T. Bhattacharya, M. Cheng, et al., 
Phys. Rev. D 80, 014504 (2009); \href{https://arxiv.org/abs/0903.4379}{arXiv:0903.4379 [hep-lat]}.
\bibitem{BORSANYI_2010} S. Borsanyi et al., 
JHEP 1011, 077 (2010); \href{https://arxiv.org/abs/1007.2580}{arXiv:1007.2580 [hep-lat]}.
\bibitem{MARTINEZ_2013} G. Martinez, ``Advances in Quark Gluon Plasma", Proceedings of the 2011 Joliot Curie School, September 12-17th 2011, La Colle sur Loup, France; \href{https://arxiv.org/abs/1304.1452}{arXiv:1304.1452 [nucl-ex]}.
\bibitem{HOTQCD_2014} T. Bhattacharya et al. (HotQCD Collaboration), 
Phys. Rev. Lett. 113, 082001 (2014); \href{https://arxiv.org/abs/1402.5175}{arXiv:1402.5175 [hep-lat]}.
\bibitem{BIGI_SANDA_2009} I. I. Bigi and A. I. Sanda, ``CP Violation", Cambridge University Press, 2009. 
\bibitem{JENKINS_2013} E. E. Jenkins, A. V. Manohar, M. Trott, 
JHEP 1309, 063 (2013); \href{https://arxiv.org/abs/1305.0017}{arXiv:1305.0017 [hep-ph]}.
\bibitem{NOVELLO_SALIM_1979} M. Novello and J. M. Salim, 
Phys. Rev. D 20, 377 (1979).
\bibitem{NOVELLO_BERGLIAFFA_2008} M. Novello and S. E. Perez Bergliaffa, 
Phys. Rept. 463, 127 (2008); \href{https://arxiv.org/abs/0802.1634}{arXiv:0802.1634 [astro-ph]}.  
\bibitem{SALOPEK_BOND_BARDEEN_1989} D. S. Salopek, J. R. Bond and J. M. Bardeen, 
Phys. Rev. D 40, 1753 (1989).
\bibitem{FAKIR_UNRUH_1990} R. Fakir and W. G. Unruh, 
Phys. Rev. D 41, 1783 (1990).
\bibitem{KAISER_1995} D. I. Kaiser, 
Phys. Rev. D 52, 4295 (1995); \href{https://arxiv.org/abs/astro-ph/9408044}{arXiv:astro-ph/9408044}.
\bibitem{KOMATSU_FUTAMASE_1999} E. Komatsu and T. Futamase, 
Phys. Rev. D 59, 064029 (1999); \href{https://arxiv.org/abs/astro-ph/9901127}{arXiv:astro-ph/9901127}.
\bibitem{BEZRUKOV_SHAPOSHNIKOV_2008} F. L. Bezrukov and M. Shaposhnikov, 
Phys. Lett. B 659, 703 (2008); \href{https://arxiv.org/abs/0710.3755}{arXiv:0710.3755 [hep-th]}.
\bibitem{BARVINSKY_KAMENSHCHIK_STAROBINSKY_2008} A. O. Barvinsky, A. Y. Kamenshchik and A. A. Starobinsky, 
JCAP 0811, 021 (2008); \href{https://arxiv.org/abs/0809.2104}{arXiv:0809.2104 [hep-ph]}.
\bibitem{PDG_QCD} S. Bethke, G. Dissertori, and G. P. Salam, ``Quantum Chromodynamics", PDG reviews, \href{http://pdg.lbl.gov}{pdg.lbl.gov} (2015).
\bibitem{FOCK_1929} V. Fock, Z. Phys. 57, 261 (1929).
\bibitem{BENDER_BOETTCHER_1998} C. M. Bender, and S. Boettcher, 
Phys. Rev. Lett. 80, 5243 (1998); \href{https://arxiv.org/abs/physics/9712001}{arXiv:physics/9712001 [math-ph]}.
\bibitem{BENDER_2007} C. M. Bender, 
Rept. Prog. Phys. 70, 947 (2007); \href{https://arxiv.org/abs/hep-th/0703096}{arXiv:hep-th/0703096}.
\bibitem{MANNHEIM_2015} P. D. Mannheim, 
(2015)	\href{https://arxiv.org/abs/1506.08432}{arXiv:1506.08432 [quant-ph]}. 
\bibitem{MANNHEIM_2016} P. D. Mannheim, 
(2016) \href{https://arxiv.org/abs/1512.04915}{arXiv:1512.04915 [hep-th]}. 
\bibitem{SCHRODINGER_1932} E. Schr\"{o}dinger, Sitz. Preuss. Akad. Wiss. Berlin 105 (1932).
\bibitem{CASTRO_2011} L. B. Castro, 
Phys. Lett. A 375, 2510 (2011); \href{https://arxiv.org/abs/1103.2460}{arXiv:1103.2460 [quant-ph]}.
\bibitem{RODINOV_2013} V. N. Rodinov, 
(2013) \href{https://arxiv.org/abs/1303.7053}{arXiv:1303.7053 [quant-ph]}.
\bibitem{NOVELLO_2011a} M. Novello, 
Class. Quantum Grav. 28, 035003 (2011).
\bibitem{NOVELLO_2012}  M.Novello, 
Phys. Rev. D 86, 063510 (2012).
\bibitem{KOGUT_STEPHANOV_2004} J. Kogut and M. A. Stephanov, ``The Phases of Quantum Chromodynamics: From Confinement to Extreme Environments", Cambridge University Press, 2004.
\bibitem{LETESSIER_RAFELSKY_2004} J. Letessier and J. Rafelsky, ``Hadrons and Quark-Gluon Plasma", Cambridge University Press, 2004.
\bibitem{SARKAR_2010} S. Sarkar, H. Satz and B. Sinha (eds.), ``The Physics of the Quark-Gluon Plasma", Springer, 2010.
\bibitem{CHODOS_1974} A. Chodos, R. L. Jaffe, K. Johnson, C. B. Thorn, and V. F. Weisskopf, 
Phys. Rev. D 9, 3471 (1974).
\bibitem{GREINER_QCD} W. Greiner, S. Schramm, and E. Stein, ``Quantum Chromodynamics", Springer, 2002.
\bibitem{SABZEVARI_2002} B. Sheikholeslami-Sabzevari, 
Phys. Rev. C 65 (2002) 054904.
\bibitem{DORING_2007} M. D\"{o}ring, K. Hubner, O. Kaczmarek, and F. Karsch, 
Phys. Rev. D 75 (2007) 054504; \href{https://arxiv.org/abs/hep-lat/0702009}{arXiv:hep-lat/0702009}.
\bibitem{LE_BELLAC_1996} M. Le Bellac, ``Thermal Field Theory", Cambridge University Press, 1996. 
Phys. Rev. D 52, 5206 (1995).
\bibitem{CHRISTIANSEN_MADSEN_1996} M. B. Christiansen and J. Madsen, 
 Phys. Rev. D 53, 5446 (1996).
\bibitem{FMA_1988} G. M. Fuller, G. J. Mathews and C. R. Alcock, Phys. Rev. D 37, 1380 (1988).
\bibitem{HAGEDORN_1965} R. Hagedorn, Suppl. Nuovo Cimento 3, 147 (1965).
\bibitem{BEDIAGA_2000} I. Bediaga, E. M. F. Curado, and J. M. de Miranda, Physica A 286,156 (2000); \href{https://arxiv.org/abs/hep-ph/9905255}{arXiv:hep-ph/9905255}. 
\bibitem{IVANENKO_KURDGELAIDZE_1965} D. D. Ivanenko and D. F. Kurdgelaidze, 
Astrophys. 1, 251 (1965).
\bibitem{IVANENKO_KURDGELAIDZE_1969} D. D. Ivanenko and D. F. Kurdgelaidze, 
Lett. Nuovo Cim. 2, 13 (1969).
\bibitem{SHAPIRO_TEUKOLSKY_2004} S. L. Shapiro and S. A. Tukolsky, ``Black Holes, White Dwarfs, and Neutron Stars: The Physics of Compact Objects", WILEY-VCH, 2004. 
\end{thebibliography}

\end{document}